# Tutorial on communication between access networks and the 5G core

Lucas Baleeiro Dominato Silveira*, Henrique Carvalho de Resende‡,
Cristiano B. Both†, Johann M. Marquez-Barja‡, Bruno Silvestre*, Kleber V. Cardoso*

*Universidade Federal de Goiás, Goiânia, GO, Brazil
{lucassilveira, brunoos, kleber}@inf.ufg.br

‡IDLab—Department of Applied Engineering, University of Antwerp—IMEC, Antwerp, Belgium
{henrique.carvalhoderesende, johann.marquez-barja}@uantwerpen.be

†Applied Computing Graduate Program, University of Vale do Rio dos Sinos, Porto Alegre, RS, Brazil
{cbboth}@unisinos.br

*Abstract*—Fifth-generation (5G) networks enable a variety of use cases that require differentiated connectivity, e.g., Ultra-Reliable and Low-Latency Communications (URLLC), enhanced Mobile Broadband (eMBB), and massive Machine Type Communication (mMTC). To explore the full potential of these use cases, it is mandatory to understand the communication along with the 5G network segments and architecture components. User Equipment (UE), Radio Access Network (RAN), and 5G Core (5GC) are the main components that support these new network concepts and paradigms. 3rd Generation Partnership Project has recently published Release 16, including the protocols used to communicate between RANs and 5GC, i.e., Non-Access Stratum (NAS) and NG Application Protocol (NGAP). The main goal of this work is to present a comprehensive tutorial about NAS and NGAP specifications using a didactic and practical approach. The tutorial describes the protocol stacks and aspects of the functionality of these protocols in 5G networks, such as authentication and identification procedures, data session establishment, and resource allocation. Moreover, we review the message flows related to these protocols in UE and Next Generation Node B (gNodeB) registration. To illustrate the concepts presented in the tutorial, we developed the my5G Tester: a 5GC tester that implements NAS and NGAP for evaluating three open-source 5GC projects using a black-box testing methodology.

## I. INTRODUCTION

The evolution of networks towards automation comes as a natural step for optimizing the usage of services and assessing the new application requirements such as Ultra-Reliable and Low-Latency Communications (URLLC), enhanced Mobile Broadband (eMBB), and massive Machine Type Communication (mMTC) [1]. 5G networks fundamentally changes the network deployment and configuration from a static to a complete dynamic perspective. However, developing the foundation for such dynamic networks requires aligning with the state-of-the-art beyond 5G networks concepts by preparing standards and technologies for future use cases and applications. Therefore, a careful study at the current standards and specifications related to protocols such as Non-Access Stratum (NAS) and NG Application Protocol (NGAP), as well as their evaluation with envisioned future deployments, is essential for the evolution to the next generation, i.e., 6G networks [2].

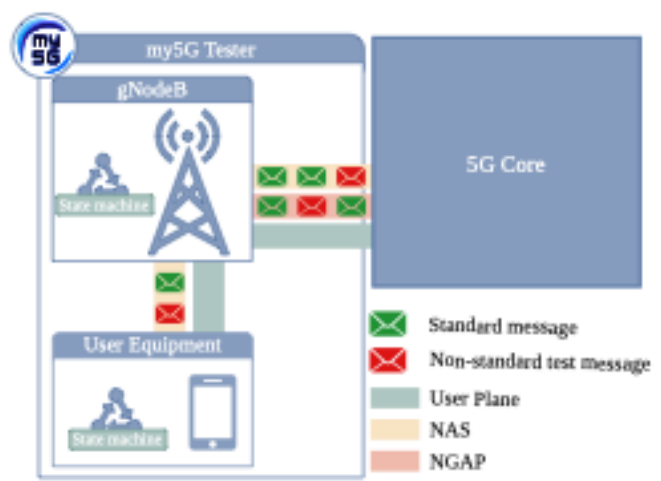

Figure 1. Overview of NAS and NGAP protocols in the 5G system.

NAS and NGAP are essential protocols for communication between 5G Core (5GC), 5G NG Radio Access Network (NG-RAN) and User Equipment (UE), as shown in Figure 1. These protocols are used for several network procedures such as registration to the network, handover, and the User Plane (UP) configuration [3]. Once NAS and NGAP enable the communication of UE to the network, these protocols are directly related to the Quality of Service (QoS) and dynamicity of the network, managing the creation of UP links and the mobility of UEs. NAS and NGAP protocols were specified by a standardization organization called the 3rd Generation Partnership Project (3GPP)[1]. The first NAS specification dates back to 2008 [4], while NGAP is an upgraded version of a protocol named S1 Application Protocol (S1AP) that dates back to 2007 [5]. Given the context in which they were designed, these protocols needed to be adapted to fit the new application requirements and the changes in the mobile network core. For instance, 5G networks are being built up around a concept called Network Slicing (NS), aiming to dynamically set virtual network layers over a shared infras-tructure to provide customized network services for different application requirements. The evolution of NAS and NGAP considered and implemented changes to empower the new dynamic aspects of 5G networks and beyond. However, with the constant evolution of research and the emergence of new application requirements [6], it is fundamental to understand these protocols specifications and regularly validate their alignment to what is being envisioned in the state-of-the-art.

### A. Motivation and Contribution

The validation of the protocols requires constant study of their specifications and implementation. However, protocols specifications are complex because, by definition, they need to deal with several details, e.g., define every message to be exchanged by the system entities given a particular scenario and state. Therefore, a tutorial explaining the NAS and NGAP protocols complements the 3GPP specifications by adding a didactic and practical approach aiming both academia and industry, supporting 5G and beyond enthusiasts to validate their ideas in these protocols.

This article provides a tutorial about NAS and NGAP protocols specifications, with the main focus on offering additional study material using a didactic and a practical approach, as abovementioned. The didactic approach would demonstrate simplified call flows on the protocols, highlighting 5G features enablers. Furthermore, the experimental process is done by implementing NAS and NGAP protocols and using them as black-box testers [7] to provide a guidance tool for mastering NAS and NGAP, as illustrated in Figure 1. Our main contributions are:

1) To provide a comprehensive tutorial on the functioning of NAS and NGAP protocols, focusing on UE and Next Generation Node B (gNodeB).
2) To detail the usage of the presented the protocol mechanics applied to the expected uses cases of the 5GC.
3) To demonstrate a Proof of Concept (POC) 5GC tester as a learning tool to reinforce the tutorial information about NAS and NGAP availing from 5GC black-box testing.

### B. Methodology

This section describes the didactic approach of this tutorial, which uses simplification or didactic quantitative reduction [8] as a learning methodology. Didactic quantitative reduction works by omitting part of the content to focus in core elements of the learning which can be afterwards extended to the omitted part of the content facilitating the learning of the whole. Therefore, we provide a strong base of knowledge to understand the protocols' specifications.

In 5G networks, NAS and NGAP protocols are of fundamental importance for registering and connecting UE and gNodeB to 5GC. Therefore, aiming at simplification, we divided this work into two parts: the theoretical approach and the practical approach. In the theoretical part, we focus on (i) registering of a new gNodeB in the infrastructure and management of NAS messages using NGAP, and (ii) a new UE registering for the first time to the network and sending Control Plane (CP) network traffic using NAS. In the practical part, we use the information introduced in the theoretical part to develop a POC tool named my5G Tester [2], demonstrating how this tutorial can be applied to a practical use case.

***Theoretical approach***: In the gNodeB i), we consider a brand new gNodeB macrocell installed in a Network Service Provider (NSP) infrastructure. The 5G NG-RAN infrastructure starts after setting up the network communication between the new macrocell and the datacenter, where 5GC is deployed. In a high-level abstraction, gNodeB will create two main links with 5GC, one for CP and another for UP. After connecting to 5GC functions, gNodeB registers to the network and enables the connection of new UE using the wireless link. At this point, NAS and NGAP protocols cooperate with each other. UE ii) connects using the wireless link to gNodeB and sends the first registration message to 5GC. All the control messages of UE with gNodeB are intermediated by gNodeB. When creating the UP communication from UE, 5GC communicates to gNodeB to create a communication channel for this UE UP. Having UP set up, UE can exchange data with Data Network (DN) and de-register from the network, followed by the de-registration of gNodeB and the de-activation of the equipment.

***Practical approach***: The study on hands-on experiments on 5GC became closer to reality with the introduction of the open-source 5GC solutions, such as Free5GC, Open5GS, and OpenAirInterface [9], [10], [11]. Nonetheless, the study on 5GC is only possible if there is an integration of the core network with real or simulated NG-RAN components. Therefore, the study on the behavior of NAS and NGAP protocols is essential to enhance the experimentation with 5GC solutions. With the knowledge of these protocols, low-cost research prototypes can be developed in order to simulate UE and gNodeB behaviors. As a practical approach for this tutorial, we rely on the theoretical use cases as a base for implementing the my5G Tester: a POC 5GC black-box tester. In order to do so, we provide a guide over the technology, architecture, and main components of the tester, explaining how it relates to NAS and NGAP protocols. We argue that the practical approach is of fundamental importance for a correct and better learning outcome. Moreover, the tester can validate the alignment of 5GC with NAS and NGAP protocols specifications, supporting the extension of the protocols for state-of-the-art 5G and beyond research.

The sections of this article are organized as follows. The 5GC main components and the NG-RAN are explained in Section II. The protocols NAS and NGAP are detailed in Section III and Section IV, respectively. In Section V, we provide the my5G Tester to deepen the NAS and NGAP protocol studies, analyzing three 5GCs open-source projects concerning conformance and robustness tests. In Section VI, we discuss open issues and challenges for 5G and beyond directly related to NAS and NGAP protocols. We present the concluding remarks and future work for this tutorial in Section VII. Last, but not least, we present the list of acronyms used along this paper in Section VIII to ease the reading.

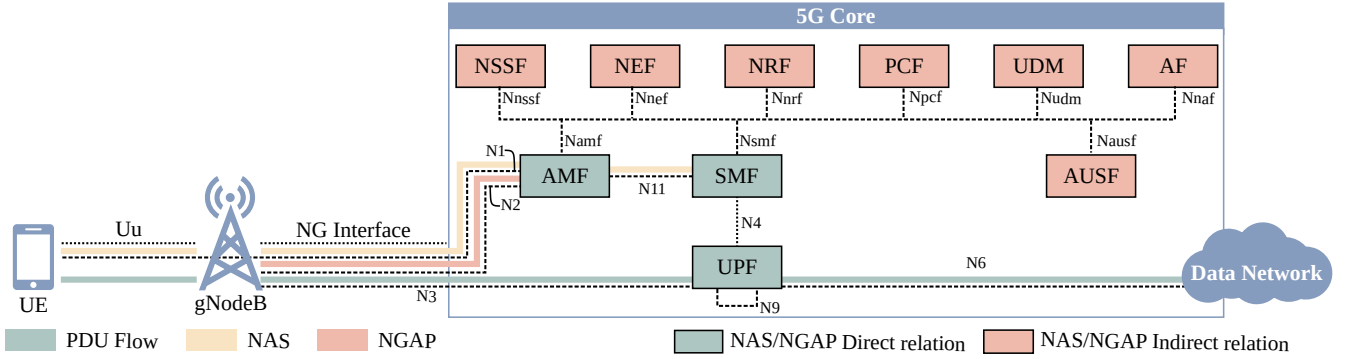

Figure 2. 5G Core (5GC) Service-based Architecture (SBA).

## II. 5G SYSTEM COMPONENTS AND INTERFACES

The 5G System (5GS) has as its main objective to provide connectivity for UEs. For this, 5GS enables the registration of UEs in the network via CP procedures and sets up the UP paths to DNs. These CP procedures require the communication of multiple components from the edge of the network to the core. 3GPP [12] designed a reference architecture to organize the communication among the 5GS components through reference points. In this context, we designed a 5GS architecture complaint by 3GPP's reference architecture to describe the functioning of NAS and NGAP protocols, as shown in Figure 2. The 5GS architecture includes two major components: NG-RAN and 5GC. UE communicates with NG-RAN over a radio interface named *Uu*, and NG-RAN communicates with 5GC over the *NG* interface. Therefore, all the communication of UE to 5GC is carried over *NG* interface. We present a brief overview of the 5GC architecture in the next section that is necessary to understand the overall functions and the NAS and the NGAP protocols.

### A. 5G Core (5GC)

5GC was initially specified in Release 15 where it introduced the novelty SBA model [13]. This architecture model promotes the high-decoupling of the code by dividing it into microservices, a technology commonly employed in cloud environments. 5GC is mainly composed of ten Network Functions (NFs) which can be seen in Figure 2. This NFs provide several services from NSs for the Network Slice Selection Function (NSSF) to analytics in the case of the Network Data Analytics Function (NWDAF). However, only Access and Mobility Function (AMF), Session Management Function (SMF), and User Plane Function (UPF) are the NFs directly related to NAS and NGAP protocols. Therefore, we focus mainly on these three 5GC NFs in this tutorial. More information about the 5GC functions can be found in Cardoso *et al.* [14].

AMF is responsible for managing all the signaling that is not specific to user data, such as mobility or security [13], and also is one of the end-points for the CP communication. SMF handles the control signaling related to user data traffic, such as session establishment [13]. The user data NAS requests coming from the UE are sent to the NG-RAN through the wireless link. From the NG-RAN to the AMF these NAS packets are encapsulated by NGAP layer and sent to the AMF in the 5GC. The SMF itself does not have a direct connection to the NG-RAN, but it has an indirect communication through the AMF functions that forward the necessary communication between the NG-RAN and the SMF. To apply the user data traffic procedures, this NF is connected to UPF over the N4 reference point using the Packet Forwarding Control Protocol (PFCP) , which is the protocol communication to send control messages to UPF. Moreover, UPF represents the handling of user data [5], connecting directly to NG-RAN over a GPRS Tunnelling Protocol (GTP)-U, i.e., a GTP tunnel for the UP traffic. This function receives connection requests from NG-RAN and creates a GTP tunnel for each UE. Thereafter, UPF receives the user data traffic related to UEs and processes it by shaping the network traffic and collecting measurements.

The NAS and NGAP protocols work over N1 and N2 reference points to the CP communication. For example, the N1 reference point is used between UE and AMF through the NAS protocol. The N2 reference point is the essential interface between NG-RAN and AMF. Moreover, we have the N11 reference point between AMF and SMF. In this case, AMF transfers the NAS messages related to a certain SMF instance over the N11 reference point. In the following, we describe an overview of NG-RAN related to NAS and NGAP protocols to have a better understanding of the main components and their interfaces.

### B. New Generation Radio Access Network (NG-RAN)

NGAP is the protocol for CP communication between a NG-RAN node, i.e., a gNodeB and the 5GC over the New Generation (NG) logical interface. The NG interface supports the separation of CP and UP. In this case, CP over the NG interface is called NG Control-plane Interface (NG-C), and UP is called NG User-plane Interface (NG-U). As illustrated in Figure 3, these two interfaces are divided into two groups: Transport Network Layer (TNL) and Radio Network Layer (RNL).

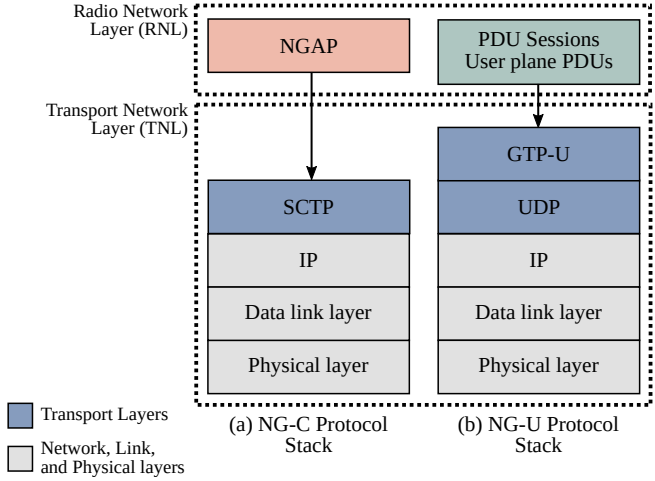

Figure 3. NG-C and NG-U protocol stack.

TNL is responsible for transmitting the network packets between the NG-RAN node and 5GC. RNL regards the control access to the mobile network. The NG-C TNL stack is implemented over Internet Protocol (IP), using Stream Control Transmission Protocol (SCTP) as the transport layer. In this case, SCTP provides a reliable channel between the NG-RAN node and AMF for signaling messages. The NG-U stack is implemented over User Datagram Protocol (UDP) using a GTP-U tunnel that provides non-guaranteed Packet Data Unit (PDU) delivery for the UP traffic flowing between the NG-RAN node and UPF. After this brief description of 5GS components and interfaces, we present in more detail NAS and NGAP in the following two sections.

## III. NON-ACCESS STRATUM (NAS)

NAS is the protocol used for control communication between UE and AMF over the N1 reference point. The communication between AMF and SMF over the N11 reference point also uses NAS, encapsulated over HTTP protocol. The UE can access the network over 3GPP networks such as gNodeB or over non-3GPP networks using technologies such as Wi-Fi or Data Over Cable Service Interface Specification (DOCSIS). In this work, we are only considering 3GPP networks. 3GPP [15] highlights the following functions of the protocols that compose NAS: (i) support for UE mobility, authentication, identification, generic UE configuration update, and security control mode procedures; (ii) support for session management procedures to establish and maintain data connectivity between UE and data network DN; and (iii) provisioning of transport for Short Message Service (SMS), LTE Positioning Protocol (LPP), Location Services (LCS), UE policy container, Steering of Roaming (SOR) transparent container, and UE parameters update information payload.

Two fundamental groups of messages support these three NAS functions: 5GS Mobility Management (5GMM) and 5GS Session Management (5GSM). 5GMM works between UE and AMF, dealing with register, mobility, security, and transport of the 5GSM protocol [16]. 5GSM is employed during the interaction between UE and SMF through AMF, offering support for connectivity management between UE and a specific DN. This connectivity management named PDU session is part of the overlay network, deployed over communication links such as the *Uu* interface and the N3 reference point. Currently, different types of PDU session are supported, e.g., IPv4, IPv6, IPv4v6, Ethernet, and unstructured. Considering the relevance of these 5GMM and 5GSM message groups for NAS, we describe their main procedures and message flows. We also provide a detailed image of the sequence of these messages in Figure 4.

### A. Procedures

The 5GMM messages support six main procedures, detailed in the following:

1) *Registration*: is responsible for informing AMF that UE wants to perform a specific type of registration, supports the registration state, and carries information relevant between UE and 5GC.
2) *Primary authentication and key agreement*: supports authentication between UE and 5GC and provides the necessary keys to security contexts and subsequent validations. There are two methods of primary authentication: one based on Extensible Authentication Protocol (EAP) and another based on 5G Authentication and Key Agreement (AKA).
3) *Identification*: is responsible for supporting specific UE identification in the 5GC. In general, 5GC may request UE to provide a specific identification such as Subscription Concealed Identifier (SUCI), International Mobile Equipment Identifier (IMEI), International Mobile station Equipment Identity and Software Version Number (IMEISV), Extended Unique Identifier (EUI)-64, or the MAC address. The latest two are examples of Permanent Equipment Identifiers (PEI).
4) *Transport*: is responsible for carrying payloads between AMF and UE. The payload can be other NAS messages, information related with UE policy containers, location services messages, SMS, etc.
5) *Security mode*: is responsible for establishing the NAS security context between UE and AMF using the key derived from the primary authentication and key agreement and supported by algorithms of ciphering and integrity.
6) *Generic UE configuration update*: is responsible for updating the UE configuration concerning access and mobility.

The 5GSM messages support only one procedure:

1) *Session management*: is responsible for authentication, authorization, establishment, modification, and release of the PDU session. Moreover, this procedure involves managing resources as networking slices, QoS, and DNs.

### B. Message flows

To comprehensively analyze the NAS protocol, we describe the flows of NAS messages for a UE registration, as illustrated in Figure 4. The messages from (1) to (8) are associated with the functions of 5GMM. The first message represents the *Registration Request* from UE to 5GC. This message carries different types of information, e.g., initial registration, mobility registration updating, periodic registration updating, and emergency registration. In this tutorial, we consider the case of the initial registration. Therefore, the UE does not have a valid context and it needs to provide a 5GS mobile identity for identification on a first registration to the network such as SUCI, or temporary identifiers, e.g., 5G Globally Unique Temporary Identifier (GUTI). Important information carried in the initial registration is the request of Local Area Data Network (LADN) and Network Slices in the form of Network Slice Selection Assistance Information (NSSAI), with two values named Slice/Service Type (SST) and an optional Slice Differentiator (SD). However, this information cannot be sent without protection, i.e., as clear-text.

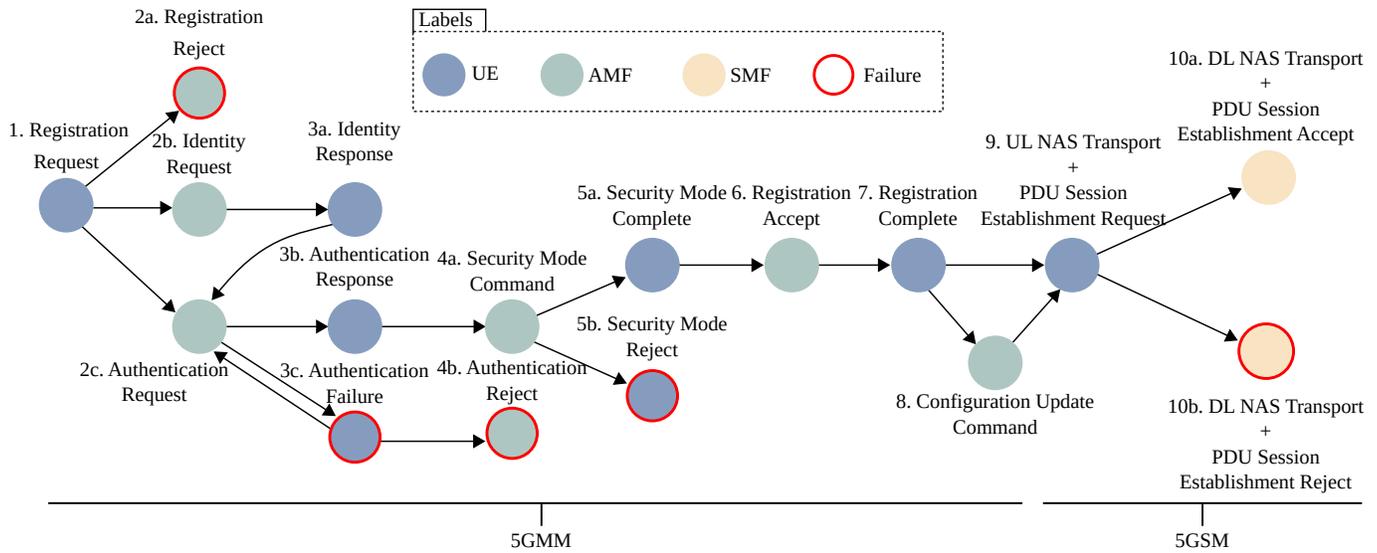

Figure 4. Non-Access Stratum (NAS) message flows.

AMF processes the *Registration Request* based on three possible messages: *Registration Reject* (2a), *Identity Request* (2b), or *Authentication Request* (2c), as shown in Figure 4. *Registration Reject* advises UE about problems in processing the *Registration Requested*, e.g., protocol errors or invalid values. *Identity Request* treats a request when UE sends an unknown identification in *Registration Request*, e.g., a 5G-GUTI unknown to AMF. In this case, 5GC triggers the *Identity Request* to ask for a specific identification to UE, including in *Identity Response* (3a) message. *Authentication Request* initiates the primary authentication and key agreement indicating that UE completed the identification. In this case, *Authentication Response* (3b) sends the answer to the authentication challenge to 5GC, which checks the value, and if the key is the same, 5GC terminates the primary authentication. Finally, UE and 5GC can create new keys for requirements such as protection of (i) NAS signaling, (ii) Radio Resource Control (RRC) signaling, (iii) user plane traffic between UE and gNodeB for 3GPP access, and (iv) Internet Key Exchange (IKE) version 2 signaling and user plane traffic between UE and Non-3GPP Interworking Function (N3WIF) for non-

3GPP access. In cases of an authentication fault, UE sends an *Authentication Failure* (3c) message allowing synchronization of Sequence Number (SQN) and a new challenge to UE. In this flow, 5GC sends *Authentication Reject* (4b) message to finish the primary authentication. The most common failures in primary authentication are related to different keys that generate problems in the verification of Message Authentication Code (MAC) and the SQN received out of range.

After exchanging messages for identification, primary authentication, and key agreement, UE and AMF establish a security context in NAS messages. The *Security Mode Command* (4a) message transports the label of selected NAS security algorithms for ciphering and integrity check performed by AMF. When UE supports the selected NAS algorithm, it responds with the *Security Mode Complete* (5a) message. UE sends the *Security Mode Reject* (5b) message if it does not support the security level. After AMF receives *Security Mode Complete*, UE has an active 5G NAS security context. Therefore, the UE can exchange encrypted messages with 5GC protecting information such as the 5GMM capability, LADN, and NSSAI requested to the 5GC. It is essential to highlight that retransmission of NAS messages can be necessary, what happens before establishment of NAS security context, such as *Registration Request* with confidential information. These messages are encapsulated in *Security Mode Complete* for forwarding to AMF. All NAS signaling must have the integrity protected and must be ciphered using the new NAS security context. The *Registration Accept* (6) message is sent to UE informing that 5GC accepts initial registration after establishing the security NAS context and authentication. This message has information such as (i) the registration area of UE, i.e., Tracking Area List (TAL), (ii) the LADN information, (iii) the list of equivalent Public Land Mobile Network (PLMN), (iv) service area restrictions, (v) allowed networking slices, (vi) timers to control periodic update registration, and (vii) the temporary identifier provided by AMF named 5G-GUTI. Finally, the *Registration Complete*

(7) message notifies AMF of the receipt of the 5G-GUTI by UE. In this stage, UE is known by 5GC regarding location, NAS connection, and security. Thus, AMF can update the UE context with the *Configuration Update Command* (8) message carrying information such as a new 5G-GUTI, TAL, service area list, LADN information, allowed or reject NSSAI, Mobile Initiated Connection Only (MICO) indication, network name, time zone, etc.

The registration messages (Request (1), Reject (2a), Accept (6), and Complete (7)) are also essential to update the 5GMM state machine inside UE and AMF. To clarify the change of states associated with the NAS message flows, we present a simplified version of the NAS 5GMM state machine in Figure 5. This figure shows the changes from the *Deregistered* state to the *Registered Initiated* state (a) after sending the *Registration Request* message into UE and when receiving the *Registration Request* message in AMF. Moreover, the 5GMM state machine presents the change from *Registered Initiated* to *Registered* (b) after receiving the *Registration Accept* message in UE and the *Registration Complete* message in AMF. It is worth to highlight that UE can request an establishment of one or more PDU sessions to 5GC in *Registered* state. Additionally, the change of states can happen from the *Registered Initiated* state to the *Deregistered* state (c) after receiving the *Registration Reject* message in UE and sending the *Registration Reject* message in AMF.

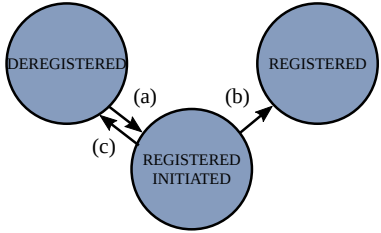

Figure 5. NAS 5GMM state machine.

The 5GSM functions are represented by messages from (9) to (10b), which focus on establishing PDU sessions, as shown in Figure 4. The *UL NAS Transport with PDU sessions Establishment Request* (9) message is sent from UE to AMF with PDU session identification, requested Single-Network Slice Selection Assistance Information (S-NSSAI), requested Data Network Name (DNN), PDU session type, among other information fields. These fields inform to 5GC requirements for the PDU session and which UPF and SMF will be selected for the PDU session. If 5GC provides the PDU session, the selected SMF sends the *PDU Session Establishment Accept* (10a) message to AMF, that encapsulates in *DL NAS transport* before sending to UE. The *PDU Session Establishment Accept* includes PDU address, QoS rules, Session Aggregate Maximum Bit Rate (AMBR), among other pieces of information. However, when SMF does not support this PDU session, it sends the *DL NAS Transport and PDU sessions Establishment Reject* (10b) message to UE, notifying the cause of rejection.

The messages that establish PDU sessions are used to update the 5GSM state machine in SMF and UE. Therefore, we show a simplified version of these states in Figure 6 to complete the analysis associated with the NAS message flows. The change from the *PDU Session Inactive* state to the *PDU Session Active Pending* state (a) occurs after sending the *PDU Session Establishment Request* message in UE and receiving the *PDU Session Establishment Request* message in SMF. The change from the *PDU Session Active Pending* state to the *PDU Session Active* state (b) happens after receiving the *PDU Session Establishment Accept* message in UE and sending the *PDU Session Establishment Accept* message in SMF. Receiving the *PDU Session Establishment Reject* message instead of the *PDU Session Establishment Accept* message changes the *PDU Session Active Pending* state to the *PDU Session Inactive* state (c) in UE. Finally, in the *PDU Session Active* state, UE actives resources to communicate with a DN using the PDU session(s) requested through 5GSM messages.

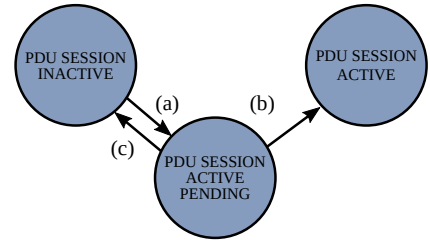

Figure 6. NAS 5GSM state machine.

## IV. NG Application Protocol (NGAP)

NGAP is the standard protocol for CP communication between the NG-RAN and 5GC which is referenced in the reference architecture by the NG-C interface and the N2 reference point. NGAP protocol works in 3GPP and non-3GPP access networks. Morevoer, NGAP supports functions such as Paging, UE Context Management, Mobility Management, PDU Session Management, NAS Transport, Warning Message Transmission, AMF Management, AMF Load Balancing, Multiple TNL Associations, Location Reporting, and UE Radio Capability Management [17]. Considering the relevance of the NGAP protocol to 5G networks, in this tutorial, we focused in the 3GPP access, describing the main procedures and message flows.

### A. Procedures

The NGAP messages support four main procedures, listed in the following:

1) *Interface management*: is responsible for maintaining the establishment of and managing the NG-C interface, where the NGAP and NAS signaling are transported. This procedure includes selecting resources as networking slices based on PLMN in AMF, managing resources as TNL connection part of the NGAP stack, and identifying between AMF and NG-RAN.

2) *Transport of NAS messages*: is responsible for carrying NAS messages between NG-RAN and AMF. NAS

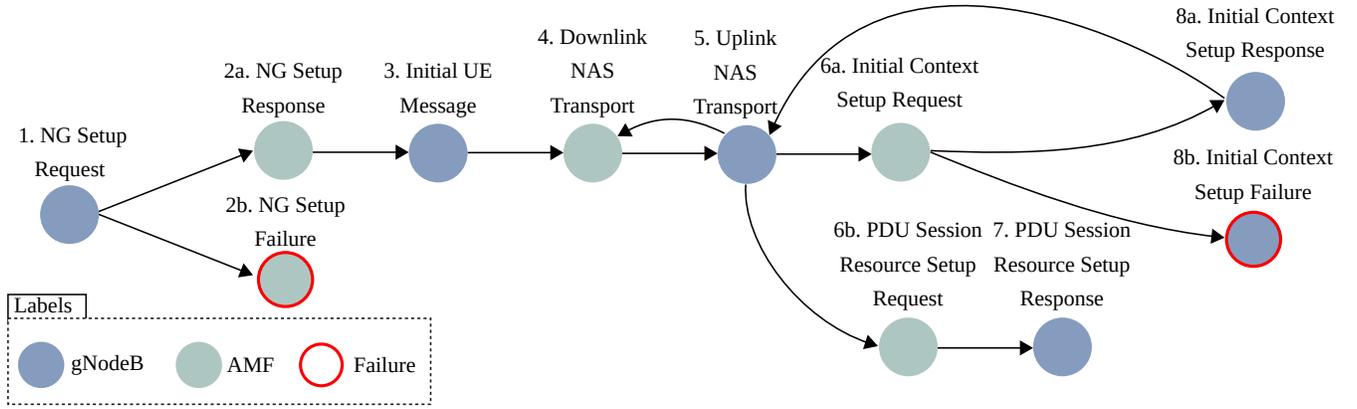

Figure 7. NG Application Protocol (NGAP) message flows.

messages are encapsulated in the NGAP protocol before forwarding to the NG-C interface.

3) *UE context management*: provides UE-based information to NG-RAN, such as security information, PDU session context, mobility restriction list, allowed networking slices, AMF connected information, UE Radio Capability, UE security capabilities, etc.

4) *PDU session management* : is responsible for managing resources necessary for determined PDU sessions. These resources are associated with NG-RAN and 5GC, establishing the Radio Interface (Uu) and NG-U interfaces for the data plane.

### B. Message flows

We describe the flows of NGAP messages for a gNodeB registration ( Figure 7 items 1, 2a, and 2b) in infrastructure and subsequently the UE registration ( Figure 7 items 3, 4, 5, 6a, 6b, 7, 8a, 8b) to comprehensively analyze the NGAP protocol. The first NGAP message sent by gNodeB represents the *NG Setup Request* and is associated with the interface management procedure. *NG Setup Request* carries information about gNodeB, e.g., Tracking Area (TA), PLMN, RAN node, Radio Access Network (RAN) Node Name, Global Node Identifier, and NSSAI supported list. These information types allow 5GC to identify the gNodeB connected to the 5G core, for example, in the Paging function, where TA selects gNodeBs for broadcast. The answer from AMF to gNodeB is *NG Setup Response* (2a) message, including information types as AMF name, AMF region ID, list of supported slices for PLMN, and relative AMF capacity. These information types can be used jointly by gNodeB. For example, based on the information of the AMF capacity and the slices for PLMN in the selection of AMF for each UE, gNodeB can balance control load among the set of AMFs. If the *NG Setup Request* information is considered incompatible by AMF, the answer is the *NG Setup Failure* (2b) message, notifying the cause of failure and temporarily ending the control flow in NG-C until AMF and gNodeB interoperate correctly. Examples of incompatibility are invalid Tracking Area Code (TAC), not-identified PLMN, and non-supported slice.

Between NG-RAN and AMF there is a state machine that we named NG-C state machine, which is updated by NGAP messages. We show the states of this machine in Figure 8 for a better analysis of the NGAP message flows. The change (a) from the *NG-C Inactive* state to the *NG-C Pending* state occurs after sending the *NG Setup Request* message in NG-RAN and receiving the *NG Setup Request* in AMF. The change (b) from the *NG-C Pending* state to the *NG-C Active* state happens after receiving the *NG Setup Response* message in NG-RAN and sending the *NG Setup Response* message in AMF. Additionally, the change (c) can occur from the *NG-C Pending* state to the *NG-C Inactive* state after receiving the *NG Setup Failure* message in NG-RAN and sending the *NG Setup Failure* message in AMF. Finally, in the *NG-C active* state, NG-RAN and AMF have a new context about each other and can manage the control flow based on their priorities.

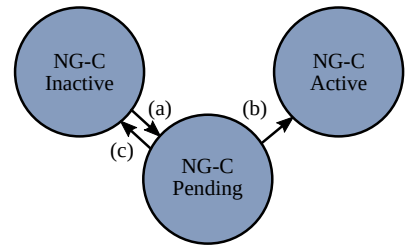

Figure 8. NGAP NG-C state machine.

Following the establishment of NG-C, gNodeB can use the NGAP protocol to transport NAS signaling to AMF. In Figure 7, the *Initial UE message* (3) is the first NGAP message to transport the UE NAS signaling to AMF. In this first case, the *Initial UE message* always carries initial NAS signaling messages, i.e, *Registration Request*. The *Downlink NAS Transport* (4) message also carries NAS messages in NG-C from AMF to NG-RAN with information about Identity Request, Authentication Request, Security Mode Command, Registration Accept, and Configuration Update Command. Similarly, the *Uplink NAS Transport* (5) subsequently carries NAS messages from NG-RAN to AMF with information about Identity Response, Authentication Response, Security Mode Complete, Registration Complete, and PDU

Session Establishment Request. Finally, in the NG-C context, it is necessary to highlight that the *Initial UE Message*, *Downlink NAS Transport*, and *Uplink NAS Transport* also support identifying specific UE NGAP traffic using identifiers provided by NG-RAN and AMF.

AMF sends the *Initial Context Setup Request* (6a) message to NG-RAN with several data: identification of the UE NGAP connection, information about AMF connected by UE, network slices allowed to UE, mobility restrictions applied to UE in cases of roaming and access, security key activated in communication for UE in RRC signaling and UP signaling, optionally information to establish one or more PDU sessions, etc. NG-RAN processes the *Initial Context Setup Request* by returning two possible messages, i.e., *Initial Context Setup Response* (8a) and *Initial Context Setup Failure* (8b), as shown in Figure 7. The *Initial Context Setup Response* notifies AMF to establish the UE context. The *Initial Context Setup Failure* notifies 5GC that NG-RAN cannot establish the UE context received in the *Initial Context Setup Request*. This failure occurs when NG-RAN does not support the established PDU session requested. These message flows finish with NG-RAN sends the *Initial Context Setup Response* message (8a), in case of success, or the *Initial Context Setup Failure* (8b) message notifying the cause of rejection to 5GC, otherwise.

In addition to the transport of NAS messages, the NGAP protocol provides for NG-RAN relevant information after the UE registration is completed in the 5GC. In order to establish the PDU session, the following messages are used: *PDU Session Resource Setup Request* (6b), *PDU Session Resource Setup Response* (7), *Initial Context Setup Request* (6a) and *Initial Context Setup Response* (8a). 5GC sends the *PDU Session Resource Setup Request* message (6b) or *Initial Context Setup Request* (6a) to NG-RAN after receiving the NAS message *PDU Establishment Request*. This message transports the necessary information to the allocation of resources in NG-RAN for one or more PDU sessions, namely: UPF Tunnel Endpoint Identifier (GTP-TEID) for creation of NG-U interface between UPF and NG-RAN, QoS information to enforce traffic of determined PDU session tied with Data Radio Bearer (DRB) part of the Uu interface, PDU session type, S-NSSAI of the PDU session, PDU session identification, the 5GSM NAS *PDU Session Establishment Accept* message, etc. Next, NG-RAN sends the *PDU Session Resource Setup Response* message (7) or *Initial Context Setup Response* (8a) message to AMF with the NG-RAN GTP-TEID to create the NG-U interface in 5GC side and information about successfully established PDU sessions. After completing this exchange of NGAP messages, the data plane path is ready, and UE can send traffic to DN using the PDU session established.

Finally, we present the simplified version of the PDU session resource state machine in Figure 9 to clarify the changes of states associated with the NGAP message flows. This state machine represents the changes from *PDU Session Resource Inactive* state to *PDU Session Resource Pending* state (a) after sending the *PDU Session Resource Setup Request* message or *Initial Context Setup Request* message from AMF and receiving *PDU Session Resource Setup Request* or *Initial Context Setup Request* in NG-RAN. NGAP messages that establish PDU session resources per UE in NG-RAN and 5GC control the states in AMF and NG-RAN. The change from *PDU Session Resource Pending* state to *PDU Session Resource Active* (b) state occurs after receiving the session management context included in *PDU Session Resource Setup Response* or *Initial Context Setup Response* message in SMF and the sending of *PDU Session Resource Setup Response* or *Initial Context Setup Response* message in NG-RAN. Additionally, the change of states can happen from the *PDU Session Resource Pending* state to *PDU Session Resource Inactive* (c) state after receiving the information about failure to setup determined PDU session. This information comes in the *PDU Session Resource Setup Response* message or in the *Initial Context Setup Failure* that reports the PDU session in the AMF. In this context, when the process is successful, the resources of the determined PDU session are allocated, moreover, NG-RAN and 5GC manage the UE traffic based on QoS rules considering the PDU session type.

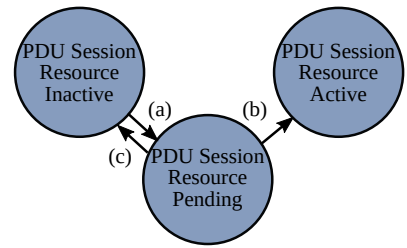

Figure 9. NGAP PDU session resource state machine.

## V. PROOF OF CONCEPT 5GC TESTER

This section presents a POC with a 5G tester called my5G Tester to deepen the NAS and NGAP protocols studies. We developed my5G Tester a tool for testing and monitoring NAS and NGAP procedures in any 5G core. Therefore, initially, we present the POC architecture details per module and explain the concepts behind the development of the POC for a better understanding of its usage. Next, we describe the experimental environment, including my5G tester, and the evaluated 5GC open-source projects, presenting all experiments and related results.

### A. POC architecture

The POC architecture is composed of four major modules, as illustrated by Figure 10. From bottom to top, the initial three modules correspond to the my5G tester[3], i.e., (i) User Interface , (ii) Controller, and (iii) Simulation. The top module corresponds to a third-party 5G core under testing as a black box. The main goal of my5G tester is to support a list of characteristics that simulate the behavior of gNodeB and UE with 5G core via NAS and NGAP protocols in order to study the communication between the access network and the 5G core. We designed these modules to have specific levels of interaction with the user, the generation of tests, and the interaction among the simulated UE, the simulated

---

[3]https://github.com/my5G/my5G-RANTester

gNodeB, and 5GC via reference points. Moreover, the my5G tester collects the data provided by this interaction for further analysis.

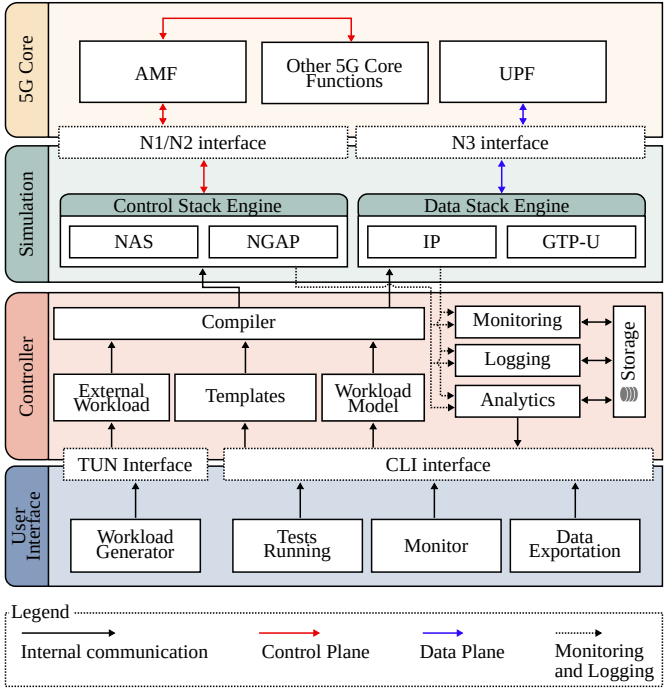

Figure 10. POC architecture.

We designed the User Interface module with two options. One option is based on Command Line Interface (CLI) which enables the experimenter to develop a script to execute a specific test with parameters and to collect output information related to the test. Moreover, CLI allows the experimenter to follow the execution of the selected test in real-time. The other option uses a virtual network interface, where the experimenter can also integrate third-party software to provided workload for testing scenarios.

The Controller module is the central entity of the POC. This module is responsible for tasks as (i) monitoring and analyzing the information provided by the simulation layer and also save this information; (ii) providing the logs about the working system; (iii) translating and processing the command or the workload received from the user interface in a test or model that can be executed in simulation layer. These tasks are executed by compiler, templates, workload model, and external workload entities.

The Simulation module provides the simulation of UE and the implementation of NAS stack to establish the N1 interface and the simulation of the NG-RAN node with the implementation of NGAP and GTP stacks, to establish the N2 and N3 interfaces. It is worth to highlight that the radio interface is not implemented between UE and NG-RAN because the main focus of my5G tester is on control and data communication between the access network and 5GC. Therefore, the radio access network and Access Stratum (AS) functionality are abstracted in order to focus on the NAS and NGAP protocols for the the 5GC black box testing.

Finally, the latest module in the top is a third-party 5G core under evaluation. The my5G tester was designed considering the input of the core under evaluation and analyzing the expected output [7]. For that reason, the test cases are based on the 3GPP specifications, which describe in detail the inputs that are supported and how they must be replied [18].

### B. Experimental scenarios

We designed the experiments using three open-source 5GC projects to analyze POC, i.e., free5GC [9], Open5GS [10], and OpenAirInterface [11]. Currently, these are the only functional 5GC projects publicly available. Furthermore, these 5G cores are analyzed using the my5G tester on version v1.0.0, sending uplink and downlink messages between the access network and 5GC using the NAS and NGAP protocols. The experiments were run in a computer with Intel Core i7@1.6GHz, 16GB RAM, and running Ubuntu 20.04 64 bit, kernel 5.4.90-050490-generic. Table I summarizes information about the experimental scenarios used in POC. Concerning OpenAirInterface (OAI) software, only the 5GC module was enabled.

| Open-source projects | Version |
|---|---|
| free5GC | v3.0.5 |
| Open5GS | v2.2.9 |
| OAI | v2021.w28 |
| my5G tester | v1.0.0 |

| Infrastructure | Configuration |
|---|---|
| RAM Memory | 16GB |
| Processor | Intel Core i7 |
| OS | Ubuntu 20.04 |
| Kernel | 5.4.90-050490-generic |

Table I. Experimental scenarios.

### C. Experimental tests

The POC is based in two groups of tests for each 5GC under evaluation. The first regards to conformance tests to verify the complete range of software testing and quality assurance requirements defined by 3GPP [19] [3]. The second group considers robustness tests using several inputs for exploring the system's behavior, e.g., error and stability of the system.

#### Conformance tests

These tests are based on sending only correct messages for the NAS procedures discussed in Section III.A and NGAP procedures presented in Section IV.A according to the 3GPP R-16 standard [19] [3]. Table II shows the eleven test procedures and set of exchanged messages for the three 5GC open-source projects. For each procedure presented in Table II, we verified the reply from each 5GC according to following the rules: (i) the state of my5G tester is compatible with the received message, (ii) the message received contains all mandatory fields, (iii) the information in the message can be understood based on its procedure, (iv) the information in the message is valid based on its procedure. We implement the machines states and flows described in Section III.B and Section IV.B inside my5G tester to achieve these conformance tests. Moreover, we analyze the Primary authentication and key agreement procedure based on the 5G-AKA method and the UE context management procedure with and without information for establishing the PDU session. All procedures tested in the three 5GC projects had the expected behavior according to the 3GPP standard, except by Generic UE configuration update. In this case, only the Open5GS project behaves as expected by sending the *Configuration Update Command* message to my5G tester in the UE registration flow.

| Procedures on testing | Protocols | Messages | free5GC | Open5GS | OAI |
|---|---|---|---|---|---|
| Registration | NAS | Registration Request, Registration Accept, and Registration Complete | ✓ | ✓ | ✓ |
| Primary authentication and key agreement | NAS | Authentication Request and Authentication Response | ✓ | ✓ | ✓ |
| Identification | NAS | Identity Request and Identity Response | ✓ | ✓ | ✓ |
| Transport | NAS | UL NAS transport and DL NAS transport | ✓ | ✓ | ✓ |
| Security mode | NAS | Security Mode Command and Security Mode Complete | ✓ | ✓ | ✓ |
| Generic UE configuration update | NAS | Configuration Update Command | ✗ | ✓ | ✗ |
| Session management | NAS | PDU Establishment Request and PDU Establishment Accept | ✓ | ✓ | ✓ |
| Interface management | NGAP | NG Setup Request and NG Setup Response | ✓ | ✓ | ✓ |
| Transport NAS messages | NGAP | Downlink NAS Transport and Uplink NAS Transport | ✓ | ✓ | ✓ |
| UE context management | NGAP | Initial Context Setup Request and Initial Context Setup Response | ✓ | ✓ | ✓ |
| PDU session management | NGAP | PDU Session Resource Setup Request and PDU Session Resource Setup Response | ✓ | ✓ | ✓ |

Table II. Conformance tests.

*Robustness tests*

We also analyze robustness tests since they are also relevant to assess the quality of the 5GC projects and their level of adherence to the 3GPP standard [19] [3]. Therefore, we defined six main robustness tests for NAS procedures discussed in Section III.A and one robustness test for NGAP procedure discussed in Section IV.A. Table III shows the tests and their results in the three evaluated 5GC projects. Additional details are presented in the following.

In the **Registration testing**, we evaluate two cases in the Registration Request message from the Registration procedure: (i) send the request without mobile identity (mandatory field), and (ii) send the request with encrypted information. In both cases, my5G tester expects AMF to return a Registration Reject message to interrupt registration flow. Only, Open5GS answered according to the expected in the two changes. The free5GC and the OAI did not send the Registration Reject message in the case (i), but interrupted the registration flow. In the case (ii), the free5GC and the OAI accepted the Registration Request with non-clear text information and persisted in the UE registration without any warning.

In the **Authentication testing**, we defined two operations to analyze the Primary authentication and key agreement procedure. First, we used invalid information in the Authentication Response message on the 5G-AKA method. We expected AMF to abort the Primary authentication and key agreement procedure, sending an Authentication Reject message in this operation. Second, we forced the situation of synchronization failure by sending of the Authentication Failure message. In this case, we expected 5GC initiates SQN re-synchronization to continue the Primary authentication and the key agreement by sending a new Authentication Request. All three 5GC projects replied to the two operations according to the expected by the 3GPP standard [19].

For the **Security testing**, we designed two incorrect configurations in the Security Mode Complete message from the Security mode procedure. In the first configuration, my5G tester receives the Security Mode Command message from AMF and replies with Security Mode Complete without the requested IMEISV information. Based on this response, it is expected from AMF a notification to my5G tester using the Registration Reject message to stop the registration process or an Identity Request asking for the IMEI information. For this robustness test, Open5GS and OAI ignored the message without IMEISV information. The free5GC project sent the Identity Request, but without the NAS security established before. In the second configuration, my5G tester receives the Security Mode Command from AMF and answers with the Security Mode Complete message without the requested re-transmission of the Registration Request information. In this case, AMF must interrupt the registration process using the Registration Reject message. OAI and free5GC ignored the message without re-transmitting the Registration Request, and Open5GS sent the reject message to abort the registration.

In the **SMF Selection testing**, we defined a change in the UL NAS Transport message from the Transport procedure. In this robustness test, my5G tester sends the UL NAS Transport message encapsulating PDU Establishment Request to 5GC, demanding a PDU session with invalid DNN and S-NSSAI information. This incorrect information does not permit the selection of an available SMF by AMF for the requested PDU session. Thus, AMF should reply to my5G tester with the DL NAS Transport message with PDU Establishment Request, informing that AMF did not forward the request to SMF (with the 5GSM message). We observed that free5GC sent the messages according to the expected. However, Open5GS and OAI did not attend SMF selection considering invalid DNN and S-NSSAI, and the test was simply interrupted (without further information).

In **UPF selection testing**, we changed the UL NAS Trans-port message from the Transport procedure. Our change makes my5G tester to send the UL NAS Transport message carrying the PDU Establishment Request to 5GC, asking for a PDU session with incorrect information: DNN and S-NSSAI invalids. This incorrect information does not permit the selection of an available UPF by SMF for the requested PDU session. In this case, 5GC must respond with the DL NAS Transport message encapsulating the PDU Establishment Reject to inform the unsuccessful operation. In this robustness test, the three 5GC projects sent the message according to the 3GPP standard [19].

| Name | Protocols | Procedures involved | Messages | free5GC | Open5GS | OAI |
|------|-----------|---------------------|----------|---------|---------|-----|
| Registration | NAS | Registration | Registration Request and Registration Reject | ✗ | ✓ | ✗ |
| Authentication | NAS | Primary authentication and key agreement | Authentication Request, Authentication Response, Authentication Failure and Authentication Reject | ✓ | ✓ | ✓ |
| Security | NAS | Security mode, Registration and Identification | Security Mode Command, Security Mode Complete, Registration Reject, Identity Request and Identity Response | ✗ | ✗ | ✗ |
| SMF selection | NAS | Transport | UL NAS Transport and DL NAS Transport | ✓ | ✗ | ✗ |
| UPF selection | NAS | Transport, Session management | UL NAS Transport, DL NAS Transport, PDU Establishment Request and PDU Establishment Reject | ✓ | ✓ | ✓ |
| NAS flow validate | NAS | Security mode, Transport, Session management and Registration | Security Mode Command, UL NAS Transport, PDU Session Establishment Request, Security Mode Complete and Registration Accept | ✓ | ✓ | ✗ |
| Interface management | NGAP | Interface Management | NG Setup Request and NG Setup Failure | ✓ | ✓ | ✗ |

Table III. Robustness tests.

In our last test with the NAS protocol, named **NAS flow validate testing**, we defined a change in the flow of NAS. In this case, my5G tester sends the UL NAS Transport message carrying the PDU Establishment Request in an incorrect order, i.e., before receiving the Registration Accept message and after receiving the Security Command Mode message. In this situation, AMF should ignore the message and re-send the Security Mode Command message after the expiry of the timer of six seconds and wait for the Security Mode Complete message instead of the UL NAS Transport message. Based on the 3GPP standard [19], this behavior must be tried four additional times, considering the expiry of the timer of six seconds on the every time. In case of failure, the registration process must be aborted, sending a Registration Reject message. In this robustness test, free5GC and Open5GS answer according to the expected by the 3GPP standard [19]. However, OAI accepted the UL NAS Transport message, including PDU Establishment Request, and replied with the DL NAS Transport message carrying PDU Establishment Accept before the Registration Accept message.

Finally, we use the NGAP protocol in the **Interface management testing** to complete our robustness tests. In this last test, my5G tester changes the NG Setup Request from the Interface Management procedure, sending invalid information for PLMN, TAC, or S-NSSAI that AMF does not support. In this context, my5G tester expects that AMF replies with the NG Setup Failure message. In this robustness testing, only OAI did not reply as expected, considering the 3GPP standard [3]. OAI AMF ignores the NG Setup Request message with invalid information associated with PLMN, e.g., sending S-NSSAI with TAC not supported by AMF, and OAI AMF does not return any error message.

This section sheds some light on the development level of the open-source 5GC projects. In summary, all of them exhibit satisfactory adherence to the standards considering a normal operation. However, even basic errors may imply an unexpected behavior in the evaluated 5GCs, especially the OAI software, which seems a less mature implementation. In the next section, we present additional insights we collected during our study in the form of a discussion about open issues.

## VI. OPEN ISSUES

Several countries around the world has started a fast deployment of 5G networks in the last years. Naturally, there is still a lot of uncertainty, and several forecasts cannot be consolidated in the first years of using this mobile generation. For example, End-to-End (E2E) slicing on 5G considering RAN, transport, and core networks is in its infancy and it still under investigation in the academy. Moreover, some topics are only basically addressed in 3GPP Releases 16, 17, and 18. For example, the Self-Organizing Networks (SON) usage can make the 5G networks more adaptive and efficient with the broad adoption of Machine Learning and Artificial Intelligence, but depends of significant volume of the information. In the following, we briefly discuss challenges for 5G and beyond directly related to NAS and NGAP protocols.

### A. Network slicing

Several contributions are being defined in 3GPP Releases 16 and 17, considering the slicing concept in the mobile network. Moreover, academia introduces new solutions and offers opportunities for network slicing research, e.g., authentication and authorization at different network slice levels [20], dynamic allocation of slices considering elasticity of computational resources [21], [22]. Through our tutorial on the functioning of NAS and NGAP protocols, we can obtain information about slice selection by UE. These protocols are responsible for authentication between UE and 5GC. Thus, it is critical to consider the NAS and NGAP protocols in addressing the challenges and solutions related to slicing in 5GC.

### B. Data-driven network

NWDAF was presented in Release 15 and improved in Releases 16 and 17, focused on collecting and analyzing data about the functionality of NFs in 5GC. The masterminds of NWDAF [23] argue that several behaviors related to UE's mobility and load, global network performance, network slice performance, data network congestion, among others, can be observed and predicted with the information stored in NWDAF. Based on the knowledge about NAS and NGAP protocols that provided of in the previous sections, we expected that a significant volume of the information present in NWDAF is related to NAS and NGAP protocols, directly or indirectly. As some examples of these increases in the data volume we can list: information about mobility, the signaling of the slicing network, activity/inactivity of UEs, activity/inactivity of NG-RAN, DNN congestion, load-balancing among NF, etc. In this context, we highlight that finding, collecting, and treating these information types carried by NAS and NGAP protocols are open issues that should be investigated in the following years.

### C. Security

Several studies [24], [25], [26] show vulnerabilities of NAS and NGAP protocols for the 5GC system, e.g., the lacking of confidentiality. Moreover, the integrity and replay protection of the NAS messages can be exposed to Man-in-the-Middle (MiTM) and Denial-of-Service (DoS) attacks. The sending of malformed packets also can trigger crashes in 5G cores related to restrictions such as buffer overflow and race conditions. These cases illustrate the challenges to maintaining the stability of the 5GC system. Therefore, testing for the 5GC system is gaining relevance with designing and implementation of tools to simulate NAS and NGAP in industry [27], [28], [29] and academia [30], [31], [32].

### D. Open RAN

Several studies [33], [34], [35] show the benefits of Open RAN environments, such as migrating some RAN functions to a cloud-native ecosystem, bringing more flexibility, interoperability, efficiency, and customization. In this context, more control and management of the RAN functions are expected to achieve these objectives. The information carried in the NAS and NGAP protocols is essential to help cooperation between 5GC and RAN to address the customers' demands. For example, we can highlight the allocation of radio resources based on requirements of the PDU session established by UE, de-allocation of radio resources based on UE's behavior, load-balancing across multiples NG-RAN, on-demand handover management, among others. All these information types should be processed and carried through the open interfaces defined by Open RAN initiatives, guaranteeing interoperability among the devices and networks.

### VII. Concluding Remarks and Future Work

In this article, we presented a comprehensive tutorial on 5GC access focusing on 3GPP access, NAS and NGAP protocols. Initially, we introduced the 5G system and its main interfaces. After, we described NAS and NGAP protocols showing their roles in 5G networks, e.g., NFs selection, slice networking selection, authentication, identification, establishing security communication, and allocation of resources. We also discussed NAS and NGAP messages related to 5G components, states machine, and message flows. Moreover, we also presented a POC with our my5G-RAN tester to illustrate the background provided by this tutorial, emulating the behavior of NAS and NGAP protocols with different implementations of 5G cores. We evaluated three open-source 5GC projects in terms of conformance and robustness, taking the 3GPP standard as reference. We argue that the information present in this tutorial helps the understating of 5G networks and contributes to advance scenarios related to beyond-5G and 6G.

As future work, we plan to evolve my5G-RAN tester and other open-source related tools in several directions, such as: (i) to enhance the NAS and NGAP protocols considering handover and paging procedures; (ii) to design and implement load tests using the NAS and NGAP protocols to emulate multiple UEs and NG-RANs connected to 5GC; (iii) to add new test cases, such as load tests with network slicing and fuzzy tests; (iv) to improve validations in conformance tests based on 3GPP Releases 16, 17, and beyond; (v) to improve system monitoring through integration with analytics and data visualization tools, e.g., Grafana and Prometheus.

### VIII. Acronyms

| | | |
|---|---|---|
| **3GPP** | 3rd Generation Partnership Project | 1 |
| **5GC** | 5G Core | 1 |
| **5GS** | 5G System | 3 |
| **5GMM** | 5GS Mobility Management | 4 |
| **5GSM** | 5GS Session Management | 4 |
| **AMF** | Access and Mobility Function | 3 |
| **AS** | Access Stratum | 9 |
| **AKA** | Authentication and Key Agreement | 4 |
| **CLI** | Command Line Interface | 9 |
| **CP** | Control Plane | 2 |
| **DN** | Data Network | 2 |
| **DNN** | Data Network Name | 6 |
| **DOCSIS** | Data Over Cable Service Interface Specification | 4 |
| **DRB** | Data Radio Bearer | 8 |
| **DoS** | Denial-of-Service | 12 |
| **eMBB** | enhanced Mobile Broadband | 1 |
| **EUI** | Extended Unique Identifier | 4 |
| **EAP** | Extensible Authentication Protocol | 4 |
| **GUTI** | Globally Unique Temporary Identifier | 5 |
| **GTP** | GPRS Tunnelling Protocol | 3 |
| **IMEI** | International Mobile Equipment Identifier | 4 |
| **IMEISV** | International Mobile station Equipment Identity and Software Version Number | 4 |
| **IKE** | Internet Key Exchange | 5 |
| **IP** | Internet Protocol | 4 |
| **LADN** | Local Area Data Network | 5 |
| **LADN** | Local Area Data Network | 5 |
| **LCS** | Location Services | 4 |
| **LPP** | LTE Positioning Protocol | 4 |

## Acknowledgment

This work is supported by the Brazilian National Council for Research and Development (CNPq) in cooperation with the my5G initiative and HAI-SCS project.